\documentstyle[aps,prl,epsf,preprint,tighten,floats]{revtex}
\begin{document}
\draft
\newcommand{\w}{\omega}
\newcommand{\W}{\Omega}
\title{FRONT PROPAGATION IN EVANESCENT MEDIA}
\author{M. B\"uttiker$^{\rm a}$ and H. Thomas$^{\rm b}$}
\address{
  $^{a}$D\'epartement de Physique Th\'eorique, Universit\'e de Gen\`eve,
        CH-1211 Gen\`eve 4, Switzerland\\
  $^{b}$Institut f\"ur Physik, Universit\"at Basel,
        CH-4056 Basel, Switzerland
}
\maketitle

\begin{abstract}
We investigate the time evolution of 
waves in evanescent media generated  
by a source within this medium and 
observed at some distance away from the location of the source.
The aim is to find a velocity which describes a causal
process and is thus, for a medium with relativistic dispersion, 
limited by the velocity of light. 
The wave function consists of a broad frequency forerunner
generated by the onset of the source, and of a
monochromatic front
which carries the oscillation frequency of the source.
For a medium with Schr\"{o}dinger-like dispersion 
the monochromatic front propagates with a velocity
which is in agreement with the traversal time,
and in the relativistic case the velocity 
of the fronts is limited by the velocity of light. 
For sources with a sharp onset, the forerunners are not
attenuated and in magnitude far exceed the monochromatic front.
In contrast, for sources which are frequency-band limited, the
forerunners are also attenuated and become comparable
to the monochromatic front: like in the propagating 
case, there exists a time at which a broad frequency forerunner
is augmented by a monochromatic wave. 
\smallskip
\end{abstract}

Fifteen years ago Rolf Landauer in collaboration with one of 
the authors\cite{MBRL82} revived an old question: What is the
time of interaction of a tunneling particle with the barrier? 
This work was motivated by the insight that the most often 
used approach which follows peaks of wave packets as they approach 
and emerge from a tunneling barrier is dubious: there is no 
causal relationship between the peak of an incident wave packet 
and the peak of an emerging wave packet. Instead a novel approach was 
needed:  Ref. 1 and subsequent work\cite{SCRIPT,TMRL93} investigated 
tunneling through a barrier with an oscillating potential and analyzed 
tunneling as a function of the oscillation frequency. 
Similarly, the Larmor clock, originally proposed by Baz\cite{BAZ} and 
Rybachenko\cite{RYBA}, 
based on the precession of the spin of particles 
was re-analyzed and the importance of spin rotation 
was pointed out\cite{LARMOR}. Both the oscillating barrier 
and the Larmor clock lead to traversal times which differ 
from a stationary phase analysis. 
We can indicate here the wide interest and 
the broad discussion which these works have generated 
only with a reference to a few recent works\cite{RECENT} and 
by a number of reviews\cite{RLTM94,HAUGE,EAVES,LEAV88}.  
 
While there have been only a few experiments on electrical 
systems\cite{EXPERI}, recently a number of interesting 
experiments
which exploit the analogy\cite{TMRL92} between electron tunneling 
and tunneling of photons\cite{OPTICS} or the propagation of classical 
microwaves in evanescent media\cite{MICRO,NIMTZ,SPIELM,DEUGOL,BALCOU} 
have provided a new impetus to the field. 
Of particular importance are the apparently superluminal 
tunneling times reported in these experiments. 
The superluminal appearance of wave packet peaks can 
be explained according to 
Deutch and Low\cite{DEUTCH} by demonstrating
that the transmitted wave packet is 
made up mainly of contributions 
which stem from the head of the incident wave packet. 
Japha and Kurizki\cite{JAPHAK} explain that 
the connection between the transmitted 
wave packet and the leading portion 
of the incident wave is a consequence of destructive interference
of causal retarded tunneling paths. 
It is clear that the wave evolution is inherently a causal 
process and that superluminal velocities arise from the use of an acausal
definition of the velocity like the comparison of 
the peaks of the incident and the transmitted wave.
These experiments thus raise the question of
whether it is possible to define a traversal time which 
represents a manifestly causal process. 
We can identify a number of different velocities 
which characterize wave propagation like the phase velocity, the group 
velocity, the signal velocity and the velocity of the head of the wave.  
It is well understood that both the phase velocity and the 
group velocity can exceed the velocity of light,
but that the signal velocity is always smaller than or at most
equal to the velocity of light and the velocity of the head of
the wave always coincides with the velocity of light.
In principle there should exist a traversal time corresponding 
to each of these possible definitions of velocity. 
It is the purpose of this work 
to discuss a traversal time which in an evanescent medium 
corresponds to a signal velocity, and which, therefore,
characterizes a causal process.  

Interest in the signal velocity of propagating waves arose from
the observation by R.~Wien that, in apparent contradiction
to special relativity theory, the phase velocity and the 
group velocity can easily exceed the velocity of light in dispersive 
media. In response A. Sommerfeld\cite{SOMME1} started a fundamental 
discussion by analyzing the propagation of wave fronts.
While the  phase velocity and the group velocity can exceed the velocity
of light without contradicting the principles of causality and relativity,
the velocity of the front and the signal velocity remain strictly
smaller than or equal to the velocity of light\cite{SOMME2}.
A detailed discussion of this approach was subsequently given
by Brillouin\cite{BRILLO}.
The method of Sommerfeld and Brillouin considers a source 
which is quiescent up to a given instant, when it is abruptly
switched on. In a medium that allows propagation, 
the sudden onset of the source leads at the observation 
point located some distance away from the source 
to small forerunners of indefinite frequency; it is followed by a wave
which oscillates with the frequency of the 
source and marks the arrival of the signal. 
For freely propagating waves (classically allowed
regions of dispersion) the forerunners are
small and the arrival of the main part is marked by a rapid increase 
in the intensity of the wave. 
The method of Sommerfeld and Brillouin was applied to evanescent 
media with a Schr\"odinger-like dispersion by Stevens\cite{STEVEN}.
Stevens calculates an arrival time of a signal. 
He did, however, not analyze the magnitude of the 
different contributions to the total wave. 
Indeed, subsequent works 
by Teranishi et al.\cite{TERANI}, Jauho and Jonson\cite{JAUHOJ}, 
Ranfagni et al.\cite{RANFAG}
have called into question 
the very existence of a front which would mark 
the arrival of the signal in an evanescent medium. 
Moretti\cite{MORETT} emphasized exact analytical solutions,
but similarly to Stevens did not discuss the magnitude of different
contributions to the evolution of the wave. 
The lack of a main part found from numerical analysis\cite{JAUHOJ,RANFAG}
was further emphasized in more recent analytical work by 
Brouard and Muga\cite{BROUAR}. 
In contrast to Stevens who predicted an arrival time of the main
part of the wave with a traversal time which is in agreement with that 
found from the oscillating barrier or the Larmor clock, 
Ranfagni et al\cite{RANFAG} and Brouard and Muga\cite{BROUAR} 
suggest that their analysis is actually in better agreement with the 
phase time.   
Since propagation occurs now in an evanescent medium, 
it is essential to investigate the amplitudes of the front 
and the forerunners in detail. Contrary to the case of 
freely propagating waves,  the evanescent medium transmits 
the high-frequency components which make up the forerunners 
with little attenuation (the forerunners propagate in 
effect freely) while the "main part" of the wave 
is evanescent and thus exponentially suppressed. 

The situation changes dramatically if instead of a source
with a sharp onset a frequency-band limited source is used.
The source which is of interest switches on gradually
but still fast compared to the traversal time.
For a source which switches on too slowly, the traversal time
cannot be resolved.
Moreover, the highest frequency must be smaller than the
threshold which permits free propagation. Thus the frequency-band
is limited both at the high end and at the low end.
For such a frequency-band limited source, all frequencies
are in the evanescent range of the dispersion. 
This has the consequence that not only the fronts which carry
the monochromatic frequency of the source, but also the forerunners
are exponentially attenuated! The forerunners may, however, still
exceed in magnitude the monochromatic fronts since the evanescent
waves of the highest frequencies of the source are attenuated less
strongly then the evanescent wave with the frequency of the source. 
Within this limitation, we encounter for the frequency-band limited
source a wave evolution which is completely analogous to the 
freely propagating case:  at the traversal time known from  
the oscillating barrier\cite{MBRL82} and  
the Larmor clock\cite{LARMOR}, a forerunner of indefinite
frequency is augmented by a nearly monochromatic wave of 
comparable amplitude. The discussion for the frequency-band
limited source is presented for the case of a particle
with Schr\"odinger-like dispersion. The discussion for a
relativistic dispersion will be presented elsewhere\cite{MBHT}.

\subsection*{Source with a sharp onset}
We want to investigate propagation of a particle field $\psi(x,t)$
into a region $x >0$ with a constant potential $V$.
The field equation---Schr\"odinger equation for a non-relativistic
particle, Klein-Gordon equation for a relativistic boson, Dirac equation
for a relativistic fermion---has plane-wave solutions
\begin{equation}
 \psi_\w(x,t) = e^{-i[\w t - ik(\w) x]} \quad (\w\, {\rm\, real}),
\end{equation}
where the wave number $k(\w)$ is determined by the dispersion relation
\begin{equation}
 \hbar\w = V + \frac{\hbar^2 k^2}{2m}
 \quad \mbox{(non-relativistic particle)},
\label{dispnon}
\end{equation}
\begin{equation}
 (\hbar\w - V)^2 = (mc^2)^2 + c^2\hbar^2 k^2
 \quad \mbox{(relativistic particle)}.
\label{disprel}
\end{equation}
$k(\w)$ is real for propagating waves and imaginary for evanescent waves.
Its sign is determined by the boundary condition that $\psi_\w(x,t)$ is
an outgoing wave in the $+x$-direction: Positive group velocity for
propagating waves and exponential decay with $+x$ for evanescent waves
yields
\begin{eqnarray}
 k(\w)\,{\rm sign}(\w) & > & 0 \quad \mbox{for propagating waves}
\label{outpro}\\
 {\rm Im} k(\w) & > & 0 \quad \mbox{for evanescent waves}.
\label{outeva}
\end{eqnarray}
For given $\w$, the dispersion relations (\ref{dispnon},\ref{disprel})
have only a single root satisfying the boundary condition
Eq.~(\ref{outpro},\ref{outeva}).

Following Sommerfeld and Brillouin \cite{SOMME2,BRILLO}
and Stevens\cite{STEVEN} we consider an
arrangement which permits the investigation of wave front propagation:
For $t < 0$, the field is everywhere zero,
\begin{equation}
 \psi(x,t) = 0 \quad (t < 0).
\end{equation}
At $t = 0$, a source located at $x = 0$ of frequency $\w_0$ and
amplitude $A$ is switched on, i.e.,
\begin{equation}
 \psi(0,t) = A(t)\, e^{-i\w_0 t} \quad {\rm with} \quad A(t) = A\,\Theta(t),
\end{equation}
where $\Theta(t)$ is the step function, or in $\w$-space,
\begin{equation}
 \hat{\psi}(0,\w) = \hat{A}(\w - \w_0)
 = \int_0^\infty\! A(t)\, e^{i(\w - \w_0)t}\, dt
 = \frac{iA}{\w - \w_0 + i0^+}\,.
\end{equation}
The solution satisfying these initial conditions is
\begin{equation}
 \hat{\psi}(x,\w) = \hat{A}(\w - \w_0)\, e^{ik(\w)x} \quad (x > 0),
\end{equation}
or in $t$-space,
\begin{equation}
 \psi(x,t) = \frac{1}{2\pi} \int_{-\infty}^{+\infty}\!
 \hat{A}(\w - \w_0)\, e^{-i[\w t - k(\w) x]}\, d\w.
\end{equation}
For the following, it is convenient to introduce a frequency $\W$
corresponding to the kinetic energy (plus rest energy in the relativistic
case) of the particle,
\begin{equation}
 \hbar\W = \hbar\w - V.
\end{equation}
Then, the solution may be written in the form
\begin{equation}
 \psi(x,t) = \frac{iA}{2\pi}\,e^{-\frac{i}{\hbar}Vt}
             \int_{-\infty}^{+\infty}\! \frac{1}{\W - \W_0 + i0^+}\,
             e^{-i\phi (\W;x,t)}\, d\W,
\label{psi}
\end{equation}
where the phase function $\phi(\W;x,t)$ is given by
\begin{equation}
 \phi(\W;x,t) = \W t - k(\W) x.
\end{equation}
The integration is along the real $\W$-axis above the pole at
$\W = \W_0 - 0^+$ and above the branch cuts of the phase function
$\phi(\W;x,t)$.
The strategy consists in shifting the path of integration into regions
of the complex $\W$-plane where
${\rm Im}\,\phi(\W;x,t) \rightarrow -\infty$ as $|\W| \rightarrow \infty$
and collecting any contributions from the pole and the branch cuts.

\subsection*{Non-relativistic particle}
\begin{figure}
\hspace{0.0\hsize}
\epsfxsize=0.37\hsize
\epsfxsize=0.53\hsize
\epsffile{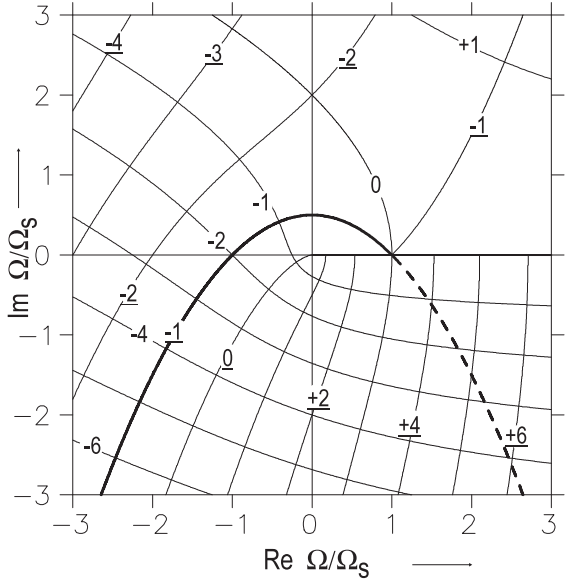}
\par

\caption{
 Form of the phase function $\phi(\W)$ in the complex $\W$-plane for
 a non-relativistic particle. Shown are the lines of constant
 $\varphi_r = {\rm Re}\,\phi/\phi_s$ for $\varphi_r = -4 \dots +6$
 (underlined numbers)
 and $\varphi_i = {\rm Im}\,\phi/\phi_s$ for $\varphi_i = -6 \dots +1$.
 The branch cut is indicated by an increased line thickness.
 The part of the line of stationary phase ${\rm Re}\,\phi = \phi_s$
 on the upper and lower sheet is shown as a full and a dashed thick
 line, respectively. After Ref. \protect\cite{MBHT}. 
 }
\label{nonrel}
\end{figure}

The phase function $\phi(\W)$ in the complex $\W$-plane is shown in
Fig.~\ref{nonrel}. It has a branch cut $\W = (0 \dots \infty)$.
The values of the function $\phi(\W)$ on the lower sheet are related
to those on the upper sheet by
\begin{equation}
 \phi^{\rm lower}(\W) = [\phi^{\rm upper}(\W^*)]^* .
\label{sheets}
\end{equation}
The phase factor $e^{-i\phi(\W)}$ has a saddle point on the real axis.
The saddle-point condition $d\phi/d\W|_s = 0$ yields
\begin{equation}
 \hbar\W_s = \frac{m}{2} \frac{x^2}{t^2}, \quad
 \hbar k_s = m\,\frac{x}{t}, \quad
 \hbar\phi_s = -\hbar\W_s t = -\frac{m}{2} \frac{x^2}{t},
\end{equation}
and the second derivative of the phase function at the saddle point is
\begin{equation}
 \phi_s^{''} = \left.\frac{d^2\phi}{d\W^2}\right|_s
             = \frac{t}{2\W_s} = \frac{\hbar}{m} \frac{t^3}{x^2}.
\end{equation}

The line $stph$ of stationary phase ${\rm Re}\,\phi(\W) = \phi_s$,
on which ${\rm Im\,\phi}$ decreases to $-\infty$ as one moves away
from the saddle point, is the parabola
\begin{equation}
 \frac{\W_i}{\W_s} = \frac{1}{2} \left[1 -
                     \left(\frac{\W_r}{\W_s}\right)^2\right],
\label{stph}
\end{equation}
where we have introduced the notation
${\rm Re}\,\W = \W_r$, ${\rm Im}\,\W = \W_i$.
On the $stph$-line, the phase function is given by
\begin{equation}
 \phi(\W) = \phi_s \left[ 1 + \frac{i}{2}
 \left( 1 - \frac{\W_r}{\W_s}\right)^2\right].
\end{equation}
For a fixed point $x$ in space, the crossing points $\W = \pm \W_s$
of the $stph$-line with the real axis move inwards with increasing $t$.
One of them crosses the pole $\W = \W_0$ at time $t = x/v_m$ where $v_m$
is a velocity given by\cite{STEVEN}
\begin{equation}
 v_m = \sqrt{\frac{2}{m}\,\hbar|\W_0|}.
\label{vf}
\end{equation}

We now deform the path of integration away from the real $\W$-axis such
that it coincides with the $stph$-line. As long as $t < x/v_m$ there is
no obstacle, and the integral remains unchanged; but for $t > x/v_m$
there occurs a loop around the pole which gives rise to a contribution
$\psi_p$ to the wave function of the form
\begin{equation}
 \psi_p(x,t) = A\,e^{-i\left(\frac{1}{\hbar}V + \W_0\right)t}\,
 e^{ik(\W_0)x}\,\Theta(v_m t - x).
\label{psip}
\end{equation}
The integral along the $stph$-line yields a contribution $\psi_s$
to the wave function,
\begin{equation}
 \psi_s(x,t) = \frac{iA}{2\pi}\,
 e^{-\frac{i}{\hbar}\left(Vt - \frac{m}{2}\frac{x^2}{t}\right)}\!
 \int_{stph} \frac{1}{\W - \W_0}\,e^{-\frac{1}{2}
 \left(1 - \frac{\W_r}{\W_s}\right)^2 \W_s t}\, d\W.
\label{psis}
\end{equation}
It may be evaluated in Gauss approximation,
\begin{equation}
 \psi_s(x,t) = \frac{iA}{2\pi}\,\sqrt{\frac{-4\pi i}{\W_s t}}\,
 \frac{\W_s}{\W_s - \W_0}\,
 e^{-\frac{i}{\hbar}\left(Vt - \frac{m}{2}\frac{x^2}{t}\right)},
\label{gauss}
\end{equation}
if the width $(\phi_s^{''})^{-1/2}$ of the exponential phase factor at the
saddle point is small compared to the distance over which the prefactor
changes significantly,
\begin{equation}
 \phi_s^{''}\, (\W_s - \W_0)^2 =
 \frac{1}{2}\left(\frac{\W_s - \W_0}{\W_s}\right)^2 \W_s t \gg 1.
\label{gausscond}
\end{equation}

The pole contribution (\ref{psip}) describes a {\it monochromatic} wave with
the frequency $\w = \w_0$ of the source, with a front which travels with
velocity $v_m$ given by Eq.~(\ref{vf}). In the propagating case
$\hbar\W_0 > 0$, $v_m$ is equal to the group velocity $v_g = d\w/dk$;
in the evanescent case $\hbar\W_0 < 0$, $v_m$ is related to the traversal
time for tunnelling\cite{MBRL82} $\tau$ for large barrier widths $L$ by
$v_m = L/\tau$.

The saddle contribution (\ref{psis}), on the other hand, describes the
perturbation due to the switching-on of the source at $t=0$ which contains
arbitrarily high frequencies; its front travels with the maximum speed
permitted by the wave equation, which is infinite for a non-relativistic
particle.

Pole contribution and saddle contribution correspond to the ``main part''
and the ``forerunner'' of Sommerfeld and Brillouin, respectively
\cite{SOMME2,BRILLO}. We will keep the expression ``forerunner'' for the
saddle contribution, but will call the pole contribution the ``monochromatic
part'' instead of ``main part'', because the latter expression is
extremely misleading in the evanescent case where the monochromatic part
is exponentially small compared to the forerunner.

It is important to point out that the existence of a wave front of the
monochromatic part $\psi_p$ does not mean that the total wave function
is discontinuous at $x = v_mt$. In fact, the solution of the Schr\"odinger
equation is continuous for all $x > 0, t > 0$. Thus, the pole contribution
$\psi_p$ must combine at $x = v_mt$ with the saddle contribution $\psi_s$
in such a way that the total wave function $\psi$ is continuous.

This continuity requirement leads to an independent determination of the
front velocity $v_m$ of the monochromatic part:
The condition that the real parts of the phases of $\psi_p$ and $\psi_s$
coincide for $x = v_mt$ reads in the propagating case
($\W_0 > 0, k(\W_0)$ real)
\begin{equation}
 \W_0t - k(\W_0)x = -\frac{m}{2\hbar}\frac{x^2}{t}
 \quad {\rm for} \quad x = v_mt,
\end{equation}
and in the evanescent case ($\W_0 < 0, k(\W_0)$ imaginary)
\begin{equation}
 -|\W_0|t = -\frac{m}{2\hbar}\frac{x^2}{t}
 \quad {\rm for} \quad x = v_mt,
\end{equation}
which yields the same result as Eq.~(\ref{vf}).
This is important because one can raise an objection against the derivation
based on the crossing of the $stph$-line with the pole: Although
the $stph$-line is the natural choice for the saddle-point integration,
this choice is by no means unique: the integral in Eq.~(\ref{psis}) does
not change if the path of integration is shifted away from the $stph$-line.

We now show explicitly that the crossing of the $stph$-line with the pole
gives rise to a discontinuity of the saddle contribution $\psi_s(x,t)$
at $x = v_mt$ which exactly compensates the discontinuity of $\psi_p(x,t)$
at the onset of the monochromatic wave.

In the propagating case, in the integral of Eq.~(\ref{psis}) we set
$\W_r = \W_s(1+u)$ where $u$ is an integration variable which measures
the distance from the saddle point $\W_s$ on the real $\W$-axis.
>From Eq.~(\ref{stph}), we obtain
\begin{equation}
 \W = \W_s \left(1 + (1-i)u - \frac{i}{2}u^2 \right); \quad
 d\W = \W_s(1 - i - iu)\,du.
\end{equation}
Then, the integral $I$ along the $stph$-line becomes
\begin{equation}
 I = \int_{-\infty}^{+\infty}\! \frac{1-i-iu}{(1-i)u-\frac{i}{2}u^2-u_0}
 \,e^{-\frac{1}{2}\alpha u^2}\,du,
\end{equation}
where we have introduced the abbreviations
\begin{equation}
 u_0 = \frac{\W_0 - \W_s}{\W_s} = \frac{(v_mt)^2 - x^2}{x^2}; \quad
 \alpha = \W_st = \frac{x^2}{2\Delta x\, v_mt}
\end{equation}
where $\Delta x = \hbar/(mv_m)$ equals 
(up to a factor of $2 \pi $) the de Broglie-wave length of a particle
moving with velocity $v_m$. The parameter in the validity condition
(\ref{gausscond}) for the Gauss approximation takes the form
\begin{equation}
 \left(\frac{\W_s - \W_0}{\W_s}\right)^2 \W_s t = \alpha u_0^2 =
 \left(\frac{v_m^{\,2}\,t^2 - x^2}{x\,v_mt}\right)^2 \frac{v_mt}{2\Delta x}.
\end{equation}
The Gauss approximation (\ref{gauss}) remains valid as long as
$\alpha u_0^2 \gg 1$, which is satisfied except very close to $x = v_mt$.
In the immediate neighborhood $x = v_mt$ where $\alpha u_0^2 \ll 1$,
on the other hand, the integral becomes
\begin{equation}
 I = \lim_{u_1\rightarrow\infty}
 \int_{-u_1}^{+u_1}\! \frac{1}{u - \frac{1}{2}(1+i)u_0}\,du =
 i\pi\,{\rm sign\,}u_0,
\end{equation}
which yields the saddle contribution
\begin{equation}
 \psi_s(x,t) = -\frac{1}{2}A\,
 e^{-\frac{i}{\hbar}\left(Vt - \frac{m}{2}\frac{x^2}{t}\right)}\,
 {\rm sign}(v_mt - x) \quad (v_mt \rightarrow x).
\end{equation}
Thus, in the propagating case, the saddle contribution has at $x = v_mt$
a jump of magnitude $A$.

In the evanescent case, the Gauss approximation (\ref{gauss}) remains
valid at $x = v_mt$ where $\W_0 = -\W_s$. However, since the wave number
$k(\W_0)$ is imaginary, the ``main part'' $\psi_p$ is exponentially small
compared to the ``forerunner'' $\psi_s$, and must therefore be compared
with the equally small contribution to the integral in Eq.~(\ref{psis})
from a narrow region around the pole. We set $\W_r = \W_s(-1+u)$ where
the integration variable $u$ measures the distance from the point $-\W_s$
on the real $\W$-axis, and $\W_0 = -\W_s(1 + w_0)$ such that
\begin{equation}
 w_0 = \frac{|\W_0| - \W_s}{\W_s} = \frac{v_m^{\,2}\,t^2 - x^2}{x^2} \quad
 {\rm whereas} \quad
 u_0 = \frac{\W_0 - \W_s}{\W_s} = -\frac{v_m^{\,2}\,t^2 + x^2}{x^2}.
\label{w0}
\end{equation}
>From Eq.~(\ref{stph}), we obtain
\begin{equation}
 \W = \W_s \left(-1 + (1+i)u - \frac{i}{2}u^2 \right); \quad
 d\W = \W_s(1 + i - iu)\,du.
\end{equation}
The contribution to the integral from the narrow region $-u_1 < u < u_1 $,
where $|w_0| \ll u_1 \ll 1, \alpha u_1 \ll 1$, is given by
\begin{equation}
 \Delta I = e^{-2\alpha}
 \int_{-u_1}^{+u_1}\! \frac{1}{u + \frac{1}{2}(1-i)w_0}\,du =
 i\pi\,e^{-2\alpha}\,{\rm sign\,}w_0
\label{DeltaI}
\end{equation}
which yields the contribution to $\psi_s$
\begin{equation}
 \Delta\psi_s(x,t) = -\frac{1}{2}A\,
 e^{-\frac{i}{\hbar}\left(Vt - \frac{m}{2}\frac{x^2}{t}\right)}
 e^{-\frac{m}{\hbar}\frac{x^2}{t}}\,
 {\rm sign}(v_mt - x) \quad (v_mt \rightarrow x).
\end{equation}
Thus, in the evanescent case, the jump of the saddle contribution
at $x = v_mt$ is exponentially small.

Comparison with Eq.~(\ref{psip}) shows that $\psi_s$ has indeed in both
cases a discontinuity opposite to that of $\psi_p$.

To summarize: In the propagating case the wave as observed at a distance
$x$ from the source grows to one half of the asymptotic amplitude A at a
time $t= x/v_m$ when the monochromatic (main) contribution of the source
sets in. In contrast, in the evanescent case the sharp onset of the source
generates a contribution to the wave which largely determines what is seen
at the observation point. The saddle point solution is of order $A$ at
$t= x/v_m$ when the exponentially small monochromatic contribution arrives.
For a Schr\"odinger-like dispersion there is no limiting velocity, the
sharp onset of the wave generates immediately a wave at the observation
point $x$. Next we investigate a medium with dispersion which contains
a limiting velocity.

\subsection*{Relativistic particle}
\begin{figure}
\hspace{0.0\hsize}
\epsfxsize=0.37\hsize
\epsfxsize=0.53\hsize
\epsffile{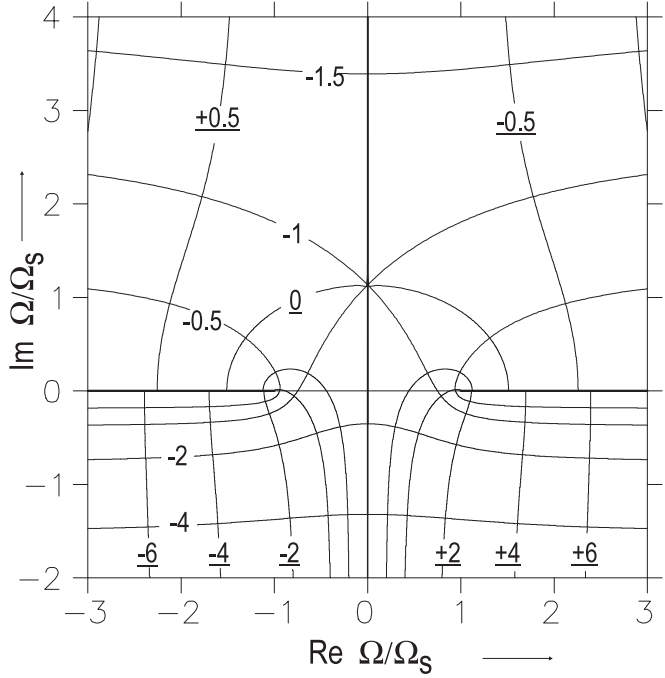}
\par

\caption{
 Form of the phase function $\phi(\W)$ in the complex $\W$-plane for
 a relativistic particle for $ct = 0.75x$. Shown are the lines of constant
 $\varphi_r = \hbar\,{\rm Re}\,\phi/(mc\protect\sqrt{x^2 - c^2t^2})$
 for $\varphi_r = -6 \dots +6\, (\W_i < 0),\; = 0, \pm 0.5\, (\W_i > 0)$
 (underlined numbers)
 and $\varphi_i = \hbar\,{\rm Im}\,\phi/(mc\protect\sqrt{x^2 - c^2t^2})$
 for $\varphi_i = -4, -2, -1.5, -1, -0.5$. After Ref. \protect\cite{MBHT}. 
}
\label{rel1}
\end{figure}

\begin{figure}

\hspace{0.0\hsize}
\epsfxsize=0.37\hsize
\epsfxsize=0.53\hsize
\epsffile{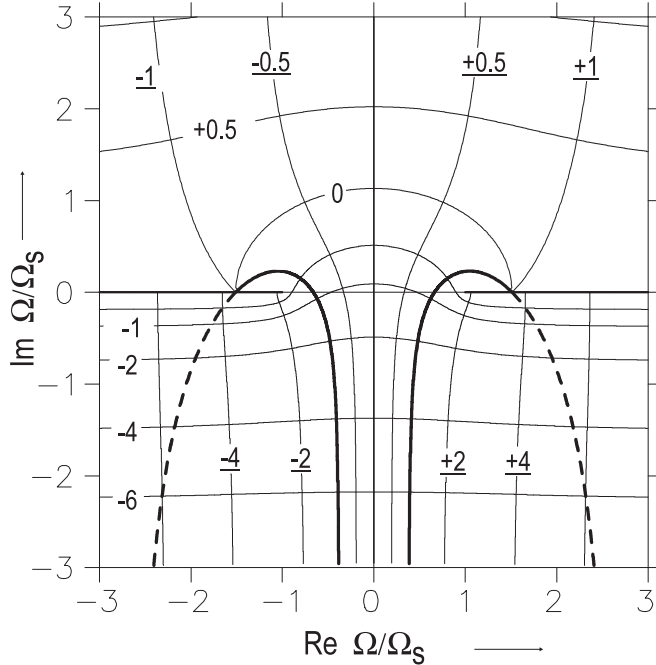}
\par 

\caption{
 Form of the phase function $\phi(\W)$ in the complex $\W$-plane for
 a relativistic particle for $ct = 1.25x$. Shown are the lines of constant
 $\varphi_r = \hbar\,{\rm Re}\,\phi/(mc\protect\sqrt{c^2t^2 - x^2})$
 for $\varphi_r = -6 \dots +6$, and
 $\varphi_i = \hbar\,{\rm Im}\,\phi/(mc\protect\sqrt{c^2t^2 - x^2})$
 for $\varphi_i = -6 \dots +1$.
 The branch cuts are indicated by an increased line thickness.
 The parts of the lines of stationary phase ${\rm Re}\,\phi = \phi_s^\pm$
 on the upper and lower sheet are shown as full and dashed thick lines,
 respectively. After Ref. \protect\cite{MBHT}. 
}
\label{rel2}
\end{figure}

The phase function $\phi(\W)$ in the complex $\W$-plane is shown in
Figs.~\ref{rel1} and \ref{rel2} for $x > ct$ and $x < ct$, respectively.
It has branch cuts
$\hbar\W = (-\infty \dots -\!mc^2)$ and $\hbar\W = (+mc^2 \dots +\!\infty)$,
and the values of $\phi(\W)$ on the upper and lower sheet are again
related by Eq.~(\ref{sheets}).

For $x > ct$, ${\rm Im}\,\phi(\W) \rightarrow -\infty$ for
${\rm Im}\,\W \rightarrow +\infty$ (see Fig.~\ref{rel1}).
By shifting the path of integration upwards, one recognizes that
\begin{equation}
 \psi(x,t) = 0 \quad \mbox{for all} \quad (x,t) \quad
 {\rm with} \quad x > ct,
\end{equation}
as required by relativistic causality.

For $x < ct$, on the other hand, ${\rm Im}\,\phi(\W) \rightarrow -\infty$
only for ${\rm Im}\,\W \rightarrow -\infty$ (see Fig.~\ref{rel2}).
In this case, the phase factor $e^{-i\phi(\W)}$ has saddle points
$\W = \pm\W_s$ on the real axis. For the saddle at $+\W_s$,
the saddle-point condition $d\phi/d\W|_s = 0$ yields
\begin{equation}
 \hbar\W_s = mc^2\,\frac{t}{\vartheta}, \quad
 \hbar k_s = m\,\frac{x}{\vartheta}, \quad
 \hbar\phi_s = \hbar\W_s\frac{\vartheta^2}{t} = mc^2\vartheta,
\label{saddlerel}
\end{equation}
where we have introduced the quantity
\begin{equation}
 \vartheta(x,t) = \sqrt{t^2 - \frac{x^2}{c^2}}.
\end{equation}
The second derivative of the phase function at the saddle $+\W_s$ is
\begin{equation}
 \phi_s^{''} = \left.\frac{d^2\phi}{d\W^2}\right|_s
             = \frac{\hbar}{m} \frac{\vartheta^3}{x^2}.
\end{equation}
The line $stph^+$ of stationary phase ${\rm Re}\,\phi(\W) = \phi_s$
through the saddle $+\W_s$, on which ${\rm Im\,\phi}$ decreases to
$-\infty$ as one moves away from the saddle point, is given by
\begin{equation}
 \W_i = -\frac{(\W_r - \W_s)[\hbar^2\W_r\W_s - (mc^2)^2]}
 {\sqrt{[\hbar^2\W_s^2 - (mc^2)^2]
 [-\hbar^2\W_r^2 + 2\hbar^2\W_r\W_s - (mc^2)^2]}}.
\label{statphaserel}
\end{equation}
It crosses the real axis at the saddle point $\W = \W_s$ and at
$\W = (mc^2/\hbar)^2/\W_s$, and goes to $\W_i = -\infty$ for
$\W_r = \W_s \pm \sqrt{\W_s^2 - (mc^2/\hbar)^2}$.
In contrast to the non-relativistic case, it is no longer possible
to obtain the imaginary part of the phase function $\phi(\W)$ on
the $stph$-lines in closed analytic form.
At the crossing point $\W = (mc^2/\hbar)^2/\W_s$, its imaginary part
has the value ${\rm Im}\,\phi = -imx^2/\hbar t$, which is the same
as the value of ${\rm Im}\,\phi$ in the nonrelativistic case at the
crossing point $\W = -\W_s$.

The results for the saddle at $-\W_s$ are obtained from
Eqs.~(\ref{saddlerel}-\ref{statphaserel}) by substituting
$\W_r \mapsto -\W_r, k_r \mapsto -k_r, \phi_r \mapsto -\phi_r$.

For a fixed point $x$ in space, the crossing points of the $stph^+$-line
with the real axis move towards $\W = mc^2$ as $t$ increases form $x/c$
to $\infty$.
In the propagating case, the saddle point $\W_s$ crosses the pole
$\W = \W_0 > mc^2/\hbar$ at time $t = x/v_m$ where $v_m$ is given by
\begin{equation}
 v_m = c\,\sqrt{1-\left(\frac{mc^2}{\hbar\W_0}\right)^2} \quad\quad
 (\hbar\W_0 > mc^2),
\label{vfrel1}
\end{equation}
which agrees with the group velocity $v_g = d\w/dk$.
In the evanescent case, the point $\W = (mc^2/\hbar)^2/(\W_s)$ crosses
the pole $\W = \W_0 < mc^2/\hbar$ at time $t = x/v_m$ where $v_m$ is
given by
\begin{equation}
 v_m = c\,\sqrt{1-\left(\frac{\hbar\W_0}{mc^2}\right)^2} \quad\quad
 (\hbar\W_0 < mc^2),
\label{vfrel2}
\end{equation}
which is equal to $\hbar |k|/m$. We relate $v_m$ to the relativistic
traversal time for tunneling $\tau_{tr}$ for large barrier widths $L$
by $v_m = L/\tau$.

\begin{figure}

\hspace{0.0\hsize}
\epsfxsize=0.37\hsize
\epsfxsize=0.53\hsize
\epsffile{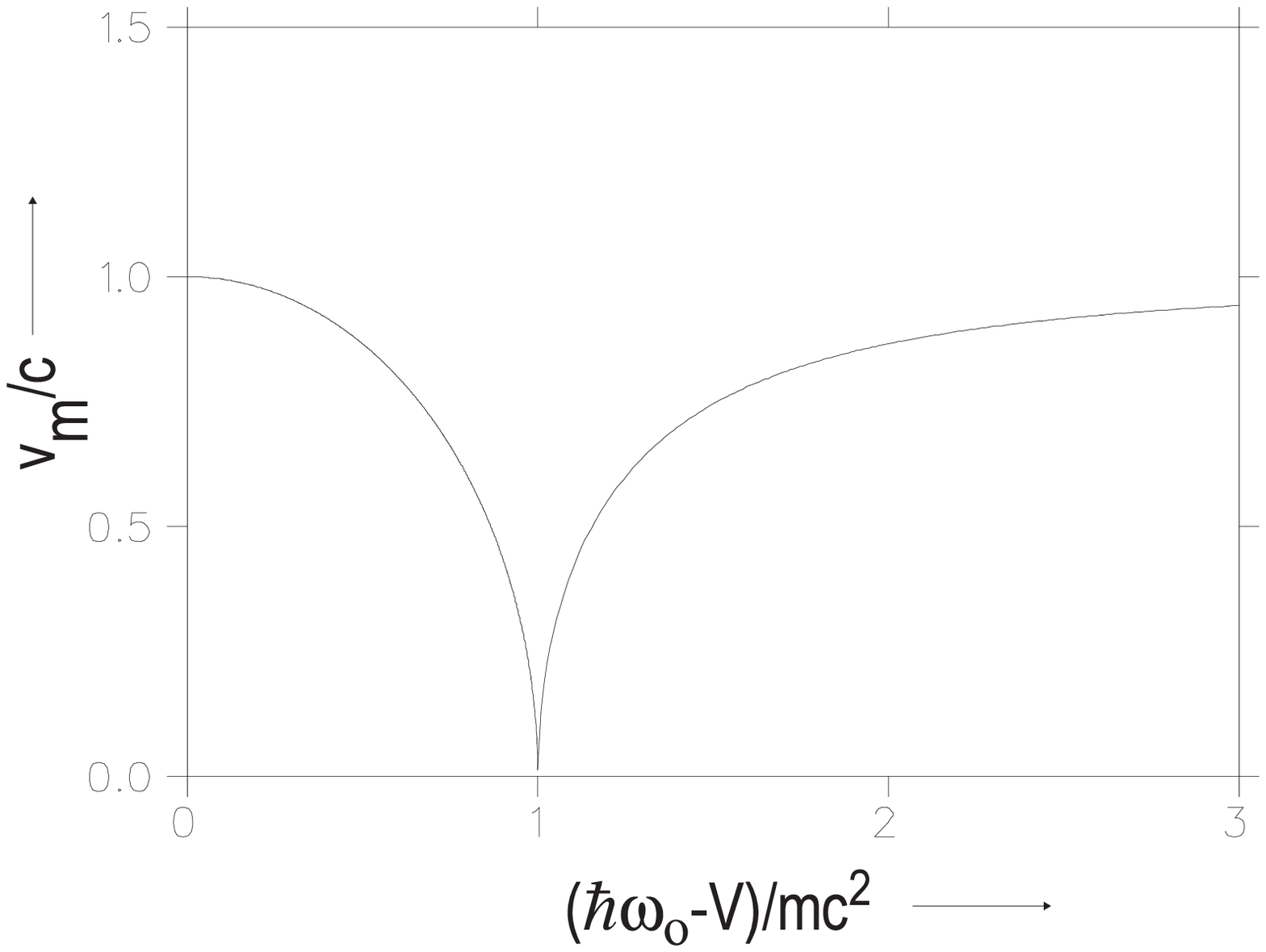}
\par

\caption{
 Front velocity of the monochromatic wave for a relativistic particle
 as function of energy. After Ref. \protect\cite{MBHT}. 
}
\label{front}
\end{figure}

The velocity $v_m$ exceeds the velocity of light $c$ neither in the
propagating nor in the evanescent case; it becomes equal to $c$ only
for zero energy and in the limit of infinite energy.
The energy dependence of $v_m$ is shown in Fig.~\ref{front}. 

In order to identify the contributions to the wave functions $\psi(x,t)$,
we proceed in the same way as in the non-relativistic case. The path
of integration is deformed away from the real $\W$-axis such that it
coincides with the $stph^\pm$-lines at positive and negative energies.
As long as $t < x/v_m$, no obstacle is encountered, and the integral
remains unchanged; but for $t > x/v_m$ there occurs a loop around the
pole which gives rise to a pole contribution
\begin{equation}
 \psi_p(x,t) = A\,e^{-i\left(\frac{1}{\hbar}V + \W_0\right)t}\,
 e^{ik(\W_0)x}\,\Theta(v_m t - x),
\label{psiprel}
\end{equation}
describing a monochromatic wave with front velo\-ci\-ty $v_m$ given by
Eqs.~(\ref{vfrel1},\ref{vfrel2}).
The integral along the $stph^\pm$-lines yield saddle contributions
\begin{equation}
 \psi_s^\pm (x,t) = \frac{iA}{2\pi}\,
 e^{-i\left(\frac{1}{\hbar}Vt \pm \phi_s \right)}\!
 \int_{stph^\pm} \frac{1}{\W - \W_0}\,e^{-{\rm Im}\,\phi_i(\W)}\, d\W
 \:\Theta(ct - x).
\label{psisrel}
\end{equation}
with a front which moves with velocity $c$.
The contribution due to the saddle at $+\W_s$ describes the excitation
of particles, and that due to the saddle at $-\W_s$ the excitation of
antiparticles. The latter part will be significant only for energies
deep in the classically forbidden region, $\hbar\W_0 \ll mc^2$.
The saddle contributions may be evaluated in Gauss approximation,
\begin{equation}
 \psi_s^\pm(x,t) = \frac{iA}{2\pi}\,
 \sqrt{\frac{\mp 2\pi i\hbar}{mc^2\vartheta}\frac{x^2}{c^2t^2}}\,
 \frac{\W_s}{\W_s \pm \W_0}\,
 e^{-\frac{i}{\hbar}\left(Vt \pm mc\sqrt{c^2t^2 - x^2}\right)}\,
 \Theta(ct - x).
\label{gaussrel}
\end{equation}
if 
\begin{equation}
 \phi_s^{''}\, (\W_s \mp \W_0)^2 =
 \frac{\hbar}{m}\frac{\vartheta^3}{x^2}\, (\W_s \mp \W_0)^2 \gg 1.
\label{gausscondrel}
\end{equation}

Since the solution of the wave equation is continuous for all $x < ct$,
the pole contribution $\psi_p$ must combine at $x = v_mt$ with the
saddle contribution $\psi_s$ in such a way that the total wave function
$\psi$ is continuous. This requirement leads again to an independent
determination of the front velocity $v_m$ of the monochromatic part:
The condition that the real parts of the phases of $\psi_p$ and $\psi_s$
coincide for $x = v_mt$ reads in the propagating case
($\hbar\W_0 > mc^2, k(\W_0)$ real)
\begin{equation}
 \hbar\left(\W_0t - k(\W_0)x\right) = mc\sqrt{c^2t^2 - x^2}
 \quad {\rm for} \quad x = v_mt,
\end{equation}
and in the evanescent case ($\hbar\W_0 < mc^2, k(\W_0)$ imaginary)
\begin{equation}
 \hbar\W_0t = mc\sqrt{c^2t^2 - x^2}
 \quad {\rm for} \quad x = v_mt,
\end{equation}
which yields the same result as Eqs.~(\ref{vfrel1},\ref{vfrel2}).

Like for a non-relativistic particle, it can be shown explicitly that the
crossing of the $stph^+$-line with the pole gives rise to a discontinuity
of the saddle contribution $\psi_s^+(x,t)$ at $x = v_mt$ which exactly
compensates the discontinuity of $\psi_p(x,t)$ at the onset of the
monochromatic wave. One obtains in the propagating case
\begin{equation}
 \psi_s^+(x,t) = -\frac{1}{2}A\,
 e^{-\frac{i}{\hbar} (Vt + mc^2\vartheta )}\,
 {\rm sign}(v_mt - x) \quad (v_mt \rightarrow x).
\end{equation}
and in the evanescent case
\begin{equation}
 \Delta\psi_s^+(x,t) = -\frac{1}{2}A\,
 e^{-\frac{i}{\hbar}(Vt + mc^2\vartheta)}
 e^{-\frac{m}{\hbar}\frac{x^2}{t}}\,
 {\rm sign}(v_mt - x) \quad (v_mt \rightarrow x).
\end{equation}

Like for the non-relativistic particle, in the evanescent case the saddle
point contribution far exceeds the monochromatic contributions to the
total wave.

\subsection*{Fronts of a frequency-band limited source}

Clearly, the high frequencies generated by a source which is switched on
instantly are highly undesirable. In this section we investigate a source
which is limited in the frequency band width. The fact that experimental
signals are often frequency-band limited has already been emphasized
\cite{BRODSK,RANFA2} (that is of course a technical and not a fundamental
limitation \cite{CHIAO}).
We restrict the discussion to the case of a non-relativistic particle.
The frequency-band limited source is described by an amplitude in Fourier
space which is the product $\hat{A}(\w - \w_0) \hat\chi(\w)$ of the
amplitude $\hat{A} = iA/(\w - \w_0 + i0^+)$ of the source with
a mathematically sharp onset and a characteristic function $\hat\chi(\w)$
which limits the range of frequencies.

For simplicity, we take
\begin{equation}
 \hat\chi (\w) = \Theta (\w - (\w_0 - \Delta \w))
               - \Theta (\w - (\w_0 + \Delta \w)),
\label{charo}
\end{equation}
where $\Theta$ is the step function, although strictly speaking this
form violates causality: As can be seen from the Fourier transform,
\begin{equation}
 \chi(t) = \frac{1}{\pi} e^{-i\w_0 t}\, \frac{\sin(\Delta\w\, t)}{t},
\label{char}
\end{equation}
the source now radiates already for $t < 0$.
The frequency band width $\Delta \w$ is chosen such that
\begin{equation}
 \frac{1}{\tau} \ll \Delta\w \ll \frac{V}{\hbar} -\w_0 = |\W_0 |.
\label{cond}
\end{equation}
Here the lower limit ensures that the onset of the source is still fast
compared to the traversal time $\tau = x/v_m $ and that the effect of
causality violation is kept mall.
The upper limit assures that all frequencies contained in the source
are in the evanescent region.

We start with the integral for the wave function given by
Eq. (\ref{psi}).
The integration path is originally along the real axis from
$\W_{-} = \W_{0} - \Delta\w$ to $\W_{+} = \W_{0} + \Delta\w $
above the pole at $\W_0$.
It is now convenient to deform the integration path in such a way
that the path first follows the line of constant imaginary phase
${\rm Im}\, \phi (\W_{-}) = \phi^{-}_{i}$ which goes through
the point $\W_{-}$, then from the point where this line
(see Fig.~\ref{nonrel}) intersects the $stph$-line follows
the line $stph$ to the point where it is intersected by the line
of constant imaginary phase
${\rm Im}\, \phi (\W_{+}) = \phi^{+}_{i}$,
and then follows this line to the point $\W_{+}$.
For $\W_0 < -\W_s$, we have to pull the integration path across
the pole, which yields a monochromatic contribution to the wave
function
\begin{equation}
 \psi_p(x,t) = A\,e^{-i\left(\frac{1}{\hbar}V + \W_0\right)t}\,
 e^{ik(\W_0)x}\,\Theta(v_m t - x),
\label{psip1}
\end{equation}
of the same form as Eq.~(\ref{psip}).

The other three contributions to the integral are evaluated for
a time interval for which $-\W_s$ is close to the pole $\W_{0}$.
Furthermore, we are interested in sources with a band width
which is small compared to $|\W_0|$.
To investigate this time window, we can therefore linearize
the phase function $\phi$ around the frequency $-\W_s$.
In terms of the variables $u$ and $v$ which measure the deviation
from $\W_s$, i.e. $\W = \W_s (-1+u+iv)$, we find for the phase function
$\phi_r = {\rm Re}\,\phi = \W_s t (-1 + u - v)$ and
$\phi_i = {\rm Im}\,\phi = \W_s t (-2 + u + v)$.
Locally, around $\W = -\W_s$ the lines of stationary phase form thus
a rectangular grid. In terms of $u$ the upper and lower frequency limits
are
\begin{equation}
u_{\pm} = \frac{\W_s - |\W_0|}{\W_s} \pm \frac{\Delta \w}{\W_s}
= - w_{0} \pm \frac{\Delta \w}{\W_s}.
\label{limits}
\end{equation}
We recall from Eq (\ref{w0}) that $w_{0} = (v^{2}_m t^{2} - x^{2})/x^{2}$.
The first part $I_1$ of the integral 
along the line $\phi^{-}_i = \W_s t (- 2 + u_{-})$
extends in $u$ from $u_{-}$ to $u_{-}/2$,
the second part $I_2$ of the integral along the $stph$-line
with $\phi_r = - \W_s t$ extends in $u$ from $u_{-}/2$ to $u_{+}/2$,
and the third integral $I_3$ along $\phi^{+}_i = \W_s t (- 2 + u_{+})$
extends in $u$ from $u_{+}/2$ to $u_{+}$.

The first (index $-$) and the third integral (index $+$) is
approximated by pulling the denominator for $u = u_\pm$ in front,
\begin{equation}
 \psi_\pm(x,t) \approx \frac{(1+i)A}{2\pi}\, \frac{1}{\Delta\w\, t}\,
 e^{-\frac{i}{\hbar}\left(Vt - \frac{m}{2}\frac{x^2}{t}\right)}\,
 e^{- \frac{m}{\hbar}\frac{x^2}{t}}\,
 e^{(-\frac{i}{2} + 1) \W_s t u_\pm}\,
 \sin\left(\W_s t \frac{u_\pm}{2}\right).
\label{int1}
\end{equation}
For $|w_0| \gg \Delta \w /|\W_0|$, which holds outside a narrow time
interval around $t = \tau$, the second integral is approximated by
pulling the denominator at $u = - w_0$ in front,
\begin{equation}
 \psi_2(x,t) \approx \frac{(1+i)A}{2\pi}\,
 \frac{1}{(\W_s - |\W_0|)t}\,
 e^{-\frac{i}{\hbar}\left(Vt - \frac{m}{2}\frac{x^2}{t}\right)}\,
 e^{- \frac{m}{\hbar}\frac{x^2}{t}}\,
 e^{(\W_s - |\W_0|)t} \sinh(\Delta\w\, t).
 \label{int2}
\end{equation} 
The key point is that all three integrals are exponentially suppressed.
For $t = \tau $ we have $\exp(-mx^2/(2\hbar\tau)) = \exp(-mv_mx/\hbar)$.
The second point to notice is that because we still have a range of
frequencies, the uppermost frequencies are least suppressed.
Thus at $t = \tau $ the upper frequencies are enhanced by a factor 
$\exp(\W_s\tau u_{+}) = \exp((mv_mx/\hbar)\Delta\w/|\W_0|)$
whereas the frequencies at the lower end of the spectrum are
additionally suppressed by a factor
$\exp(\W_s\tau u_{-}) = \exp(-(mv_mx/\hbar)\Delta\w/|\W_0|)$.
Clearly the different exponential suppression of these
frequencies is unavoidable. 

Let us next discuss what happens in the narrow time interval around
$t = \tau$ where $|w_0| \ll \Delta \w /|\W_0|$, and where the expression
for the wave function  $\psi_2(x,t)$ obtained above is not valid.
For this integral, we have to reconsider a discussion analogous to
that which leads to Eq. (34). For $w_0 \neq 0$, by pulling
the exponential factor at the upper limit in front, we find
\begin{equation}
 \psi_{s,2} (x,t) \approx \frac{A}{\pi}\,
 e^{-\frac{i}{\hbar}\left(Vt - \frac{m}{2}\frac{x^2}{t}\right)}\,
 e^{- \frac{m}{\hbar}\frac{x^2}{t}}\,
 e^{\Delta\w\, t}
 \arctan\left(\frac{\Delta \w}{\W_s - |\W_0|}\right).
\label{back1}
\end{equation}
At $w_0 = 0$, on the other hand, the integral has a discontinuity
which exactly compensates the jump of the pole contribution $\psi_p$.

The discussion presented here does not show explicitly the short-time
and the long-time behavior of the stationary phase solution.
This restriction is due to the linearization of the phase function. 
If we are sufficiently far away from the point $\W_0 = - \W_s$,
we can estimate the contribution from the saddle-point solution near
the intersection point of the $stph$-line with the vertical line
that runs through $\W_{+}$.
This leads to a wave function whose magnitude is governed by
$\exp(-\frac{1}{2}\W_s t (1 - (\W_{+}/\W_s)^2))$.
For times which are short compared to $\tau$, the saddle-point
frequency $\W_s$ is large compared to $|\W_{+}|$, and the
exponential can be approximated by $\exp (-\frac{1}{2}{\W_s} t)$.
For times which are large compared to $\tau$, we have   
$\W_s \ll |\W_{+}|$, and the exponential can be approximated by
$\exp(-\frac{1}{2}{\W_s} t (\W_{+}/\W_s)^2)$
which decays with time as $\exp (-{\rm const.}\, t^{3})$.
Thus the saddle point contribution is small both at times which
are short compared to $\tau$ and at times which are long
compared to $\tau$.

To summarize: By limiting the frequency band width of the source 
we obtain a wave evolution in the evanescent medium which shares 
the essential properties of the wave evolution in the propagating case:
At a time $t = \tau $ a forerunner with a broad frequency distribution 
is augmented by a monochromatic front with the frequency of the source.

\subsection*{Discussion}

In this work we have focused on the question of whether it is possible
to find a traversal time which is associated with a causal process.
The process investigated is the propagation of monochromatic fronts.
For the case of a source with a sharp onset these monochromatic fronts
are exponentially small compared to the forerunner which is not attenuated.
However, as we have shown, if the source is frequency-band limited the
forerunners are also exponentially attenuated. For the evanescent case,
a wave evolution is found, which as in the propagating case, exhibits at
$t = \tau$ a crossover from a wave with a broad frequency spectrum
(forerunner) to a nearly monochromatic wave. We also emphasize that
the total wave function, respectively, the resulting probability
distribution is not the only observable. Possibly, an other way to make
the monochromatic fronts visible is to incorporate a detector which
is sensitive only to a narrow range of frequencies in an interval around
the main frequency $\w_0$ of the source.
Or, we could consider particles with a spin in an evanescent medium with
a weak magnetic field and could investigate the rotation of the spin as
a function of time similarly to the stationary analysis of the Larmor
clock\cite{LARMOR}. Our work shows that the investigation of the time
evolution of particle fields is a very intersting avenue of research.
The comparison of such investigations with the results from the oscillating
barrier approach and quantum clocks \cite{MBRL82,LARMOR,RLTM94} should be
particularly illustrative. Furthermore, we hope that this work stimulates
also experiments which aim to identify the monochromatic fronts discussed
in this work.

\subsection*{Acknowledgements}
M. B. has been a collaborator of Rolf Landauer for a number of years 
and has profited from his insights and stimulations much more than 
can be expressed in this brief and formal publication. 
H. T. has profited from a multitude of stimulating interactions 
with Rolf Landauer over a period of many years. 
We have benefited from discussions with P. Wittwer, 
and we thank T. Gyalog for help with the figures.

\end{document}